\documentclass[twocolumn,dvipsnames]{aastex63}

\usepackage{amsmath}
\usepackage{amssymb}
\usepackage{bm}
\usepackage{textgreek}
\usepackage[version=4]{mhchem}

\shorttitle{Ice-Coated Pebble Drift}
\shortauthors{Price, Cleeves, Bodewits, \& \"Oberg}

\newcommand\tsup[1]{\textsuperscript{#1}}

\newcommand\supn[1]{\tsup{\ensuremath{#1}}}
\newcommand\expn[2]{\ensuremath{#1 \times 10^{#2}}}

\makeatletter
\newcommand\DeclareUnit[2]{%
    \@namedef{#1}{\@ifnextchar[{\csname @with@#1\endcsname}{\csname @without@#1\endcsname}}%
    \@namedef{@with@#1}[##1]{\text{#2\supn{##1}}}%
    \@namedef{@without@#1}{\text{#2}}%
}%
\makeatother

\DeclareUnit{cm}{\text{cm}}
\DeclareUnit{mm}{\text{mm}}
\DeclareUnit{um}{\text{\textmu m}}
\DeclareUnit{s}{\text{s}}
\DeclareUnit{g}{\text{g}}
\DeclareUnit{au}{\text{au}}
\DeclareUnit{yr}{\text{yr}}
\DeclareUnit{Myr}{\text{Myr}}
\DeclareUnit{K}{\text{K}}

\newcommand\diff{\mathop{}\!\mathrm{d}}
\newcommand\St{\ensuremath{\mathrm{St}}}

\newcommand\revisionone[1]{#1}

\begin{document}

\title{Ice-Coated Pebble Drift as a Possible Explanation for Peculiar Cometary \ce{CO}/\ce{H2O} Ratios}

\author[0000-0002-3286-3543]{Ellen M. Price}
\affiliation{Center for Astrophysics~$|$~Harvard \& Smithsonian, 60 Garden St., Cambridge, MA 02138, USA}

\author[0000-0003-2076-8001]{L. Ilsedore Cleeves}
\affiliation{University of Virginia, Department of Astronomy, 530 McCormick Rd., Charlottesville, VA 22904, USA}

\author[0000-0002-2668-7248]{Dennis Bodewits}
\affiliation{Physics Department, Auburn University, Auburn, AL 36832, USA}

\author[0000-0001-8798-1347]{Karin I. \"Oberg}
\affiliation{Center for Astrophysics~$|$~Harvard \& Smithsonian, 60 Garden St., Cambridge, MA 02138, USA}

\begin{abstract}
    To date, at least three comets --- \revisionone{2I/Borisov, C/2016 R2 (PanSTARRS), and C/2009 P1 (Garradd)} --- have been observed to have unusually high \ce{CO} concentrations compared to water. We attempt to explain these observations by modeling the effect of drifting solid (ice and dust) material on the ice compositions in protoplanetary disks. We find that, independent of the exact disk model parameters, we always obtain a region of enhanced ice-phase \ce{CO}/\ce{H2O} that spreads out in radius over time. The inner edge of this feature coincides with the \ce{CO} snowline. \revisionone{Almost every} model achieves at least \ce{CO}/\ce{H2O} of unity, \revisionone{and one model reaches a} \ce{CO}/\ce{H2O} ratio $> 10$. After running our simulations for 1~\Myr, \revisionone{an average of} 40\% of the disk ice mass contains more \ce{CO} than \ce{H2O} ice. In light of this, a population of \ce{CO}-ice \revisionone{enhanced} planetesimals are likely to generally form in the outer regions of disks, and we speculate that the aforementioned \ce{CO}-rich comets may be \revisionone{more} common, both in our own Solar System and in extrasolar systems, \revisionone{than previously expected}.
\end{abstract}


\section{Introduction}


Comets provide a unique window onto the ice-phase chemistry of a protoplanetary disk. These frozen remnants are generally considered to be the most pristine record available for understanding disk midplanes' compositions. The chemical species present in solar system comets and their relative abundances provide unique and detailed insights into the protoplanetary disk that formed our planetary system \citep{MummaCharnley2011AnnuRev,Altwegg+2003SSR}, and, more recently with the discoveries of passing extrasolar comets \citep{strom2020}, other planetary systems.

Water is typically the most abundant volatile species in cometary nuclei, with carbon monoxide comprising between about 0.2\% to 23\% relative to water, with a typical value around 4\% \citep{BockeleeMorvan&Biver2017}. However, at least three notable exceptions have been observed. Specifically, the interstellar comet 2I/Borisov was measured to have \ce{CO}/\ce{H2O} between 35\% and 173\% \citep{Cordiner+2020NatAs,Bodewits+2020Nature}, significantly higher than the average cometary values for the Solar System. \citet{Bodewits+2020Nature} suggest 2I/Borisov's composition could be explained by an unusual formation environment beyond the \ce{CO} snowline, and, statistically, it is more likely that 2I/Borisov is a typical comet for its system. However, given the ubiquity of water in interstellar clouds \citep{Boogert+2015AnnuRev}, it would be challenging to have a scenario with \ce{CO} ice freezing out without abundant water ice, which has a higher binding energy than \ce{CO}. 
\revisionone{At least two additional comets, C/2009 P1 (Garradd)} and C/2016 R2 (PanSTARRS), which originate in our own solar system, have high \ce{CO} abundances as well: \revisionone{C/2009 P1 (Garradd) has a \ce{CO} production rate of 63\% that of water \citep{Feaga+2014AJ}, and C/2016 R2 (PanSTARRS)} has an even higher \ce{CO} production rate than 2I/Borisov \citep{Biver+2018A&A,McKay+2019AJ}. \revisionone{Although these comets represent a very small fraction of the comets for which \ce{CO}/\ce{H2O} has been measured, they are more difficult to explain.} Therefore we need a mechanism that can both create enhanced \ce{CO} to \ce{H2O} ratios compared to interstellar or disk-averaged \ce{CO}/\ce{H2O} abundance ratios \textit{and} create a spread of \ce{CO} to \ce{H2O} within a disk like our solar nebula.

What kinds of mechanisms could explain these unusual compositions both interior and exterior to our solar system? \citet{Biver+2018A&A} suggest that C/2016 R2 could be a piece of a differentiated comet; \revisionone{\citet{Cordiner+2020NatAs} suggest the same for 2I/Borisov.} \citet{DeSanctis+2001AJ} found that \ce{CO} and other volatiles could almost be completely absent in the upper layers of a hypothetical differentiated comet; in this scenario, 2I/Borisov and C/2016 R2 could be pieces of the cores of such differentiated comets. Alternatively, the chemistry of the planet-forming disk could evolve over time to create exotic compositions at different disk locations. For example, \citet{Eistrup+2019A&A} consider several comets and attempt to reproduce their molecular abundances with a model protoplanetary disk. Their disk model produces a maximum \ce{CO}/\ce{H2O} ratio of about $1\%$ over a range of \revisionone{disk radii from $15$~\au\ to $30$~\au\ from the central star}. However, to reproduce a comet like 2I/Borisov, we would require a ratio that could be as high as $100\%$.

In recent years, dust transport, especially radial drift, has been found to be an important factor in shaping the solid mass distribution in disks \citep{testippvi,Piso+2015ApJ,Oberg+2016ApJL,Cridland+2017ApJ}. If the timing of volatile freeze out and dust transport due to, e.g., drift, are not synced, it could become possible to create a variety of ice compositions purely due to dust dynamics.


In this paper, we explore whether a comet such as 2I/Borisov or C/2016 R2 (PanSTARRS) could form in a pocket of \ce{CO}-rich material in an otherwise \ce{H2O}-rich disk as a result of dust transport, and under what conditions such pockets could form. The paper is structured as follows. In Section~\ref{sec:methods}, we explain the equations and software used to define our disk model. Section~\ref{sec:results} presents our calculated \ce{CO}/\ce{H2O} ice ratios across a generic protoplanetary disk. We discuss the implications of these results in light of the recent findings of \revisionone{comets} and an exo-comet with high \ce{CO} abundance in Section~\ref{sec:discussion} and conclude in Section~\ref{sec:conclusions}.

\revisionone{The authors note that a similar paper \citep{Mousis2021arXiv} appeared independently during the review process for this paper.}

\section{Methods}
\label{sec:methods}

Our goal is to globally simulate the surface densities of solids and gas in a protoplanetary disk, incorporating simple adsorption and desorption processes for the chemical species we consider, \ce{H2O} and \ce{CO}. We build on the physical models of disk gas and dust following \citet{LyndenBell&Pringle1974MNRAS} and \citet{Birnstiel+2010A&A}. In addition, we take into account the time evolving disk temperature due to the pre-main sequence stellar evolution over the time scales of our model simulation. The following sections detail these model components.

\subsection{Gas dynamics}

To model the dynamics of the gas bulk \revisionone{(defined as the bulk hydrogen gas, which experiences no source terms)}, we follow \citet{LyndenBell&Pringle1974MNRAS}, which is based on the $\alpha$-disk model of \citet{ShakuraSunyaev1973A&A}. Thus, we have the partial differential equation
\begin{equation}
    \frac{\partial \Sigma_\mathrm{gas}}{\partial t} - \frac{3}{R} \frac{\partial}{\partial R} \left[R^{1/2} \frac{\partial}{\partial R} \left(\nu \Sigma_\mathrm{gas} R^{1/2}\right)\right] = 0
    \label{eqn:pringle}
\end{equation}
in the absence of sources and sinks, where $\Sigma_\mathrm{gas} \equiv \int \rho_\mathrm{gas} \diff z$ is the surface density of gas, $\nu$ is the viscosity, $R$ is the distance from the star in the $x$-$y$ plane, and $t$ is time. Viscosity is, in turn, given by
\begin{equation}
    \nu = \alpha c_s^2 / \Omega
    \label{eqn:nu}
\end{equation}
where $\alpha$ is a small parameter of our choosing, typically set to $10^{-4}$ to $10^{-2}$; $c_s$ is the local sound speed, given by
\begin{equation}
    c_s = \sqrt{\frac{k_B T}{\mu m_p}},
    \label{eqn:soundspeed}
\end{equation}
where $k_B$ is Boltzmann's constant, $T$ is the local temperature, $\mu$ is the mean molecular weight, and $m_p$ is the proton mass; and $\Omega$ is the Keplerian angular frequency,
\begin{equation}
    \Omega = \sqrt{\frac{G M_\star}{R^3}},
    \label{eqn:keplerian}
\end{equation}
with $G$ the gravitational constant and $M_\star$ the central stellar mass. Equations~\ref{eqn:pringle}, \ref{eqn:nu}, \ref{eqn:soundspeed}, and \ref{eqn:keplerian} completely define the model of the gas bulk given parameters $\mu$, $M_\star$, and $\alpha$; the local temperature field $T = T\!\left(R, t\right)$ (see Section~\ref{sec:temperature}; and the initial condition $\Sigma_\mathrm{gas}\!\left(R, t = 0\right)$.

For the initial condition, we first define the self-similar solution,
\begin{equation}
    \Sigma_\mathrm{ss}\!\left(R\right) = \Sigma_c \left(\frac{R}{R_c}\right)^{-\gamma} \exp\!\left[-\left(\frac{R}{R_c}\right)^{2 - \gamma}\right],
    \label{eqn:selfsimilar}
\end{equation}
with $\Sigma_c = 20~\g~\cm[-2]$, $R_c = 20~\au$, and $\gamma = 0.5$; for reference, \citet{Andrews+2012ApJ} uses $-1 \leq \gamma \leq 1$. \revisionone{Here, $\Sigma_c$ is the surface density at radius $R_c$, and $\gamma$ determines the slope of the power law part of the solution.} Unfortunately, when $R \ll R_c$, this solution begins to blow up, which makes it computationally difficult to handle. We use a smooth interpolation between the self-similar solution and a flat, constant surface density profile, given by
\begin{equation}
    \Sigma_\mathrm{gas}\!\left(R, t = 0\right) = \left(\Sigma_\mathrm{ss}\!\left(R\right)^{-p} + \Sigma_\mathrm{ss}\!\left(R_\mathrm{trans}\right)^{-p}\right)^{-1 / p}
    \label{eqn:surfdens}
\end{equation}
as our initial condition. We take $p = 5$ and $R_\mathrm{trans} = 1~\au$ so that the transition occurs close to the interior of the domain and the transition from the self-similar to the flat profile is not too sharp.

Though we have chosen to work in one dimension, some quantities depend on the local density $\rho$ rather than the surface density $\Sigma$. In these cases, we assume a vertical Gaussian distribution of material,
\begin{equation}
    \rho\!\left(R, z\right) = \frac{\Sigma\!\left(R\right)}{\sqrt{2 \pi} h_\mathrm{gas}} \exp\!\left[-\frac{1}{2} \left(\frac{z}{h_\mathrm{gas}}\right)^2\right]
\end{equation}
where the scale height $h_\mathrm{gas} = c_s / \Omega$.

\subsection{Dust dynamics}

We consider two solid populations in our model: a small ``dust'' population with radius $0.1$~\um\ and a ``pebble'' population with radius $1$~\mm, with mass ratios $90\%$ and $10\%$, respectively. Following \citet{Birnstiel+2010A&A}, we define the surface density evolution for each population by the partial differential equation
\begin{equation}
    \frac{\partial \Sigma_\mathrm{solid}}{\partial t} + \frac{1}{R} \frac{\partial}{\partial R} \left(R F_\mathrm{tot}\right) = 0
    \label{eqn:birnstiel}
\end{equation}
in the absence of sources and sinks, where $\Sigma_\mathrm{solid}$ is the \revisionone{solid} surface density for a single population and $F_\mathrm{tot} \equiv F_\mathrm{adv} + F_\mathrm{diff}$ is the total flux, with contributions from an advective and diffusive part. The fluxes are given by
\begin{equation}
    F_\mathrm{adv} = \Sigma_\mathrm{solid} u_\mathrm{solid}
\end{equation}
and
\begin{equation}
    F_\mathrm{diff} = -\frac{\nu}{\St^2 + 1} \frac{\partial}{\partial R} \left(\frac{\Sigma_\mathrm{solid}}{\Sigma_\mathrm{gas}}\right) \Sigma_\mathrm{gas}.
\end{equation}
In the above equations, the Stokes number is given by
\begin{equation}
    \St = \frac{\pi}{2} \frac{a_\mathrm{gr} \rho_\mathrm{gr}}{\Sigma_\mathrm{gas}}
\end{equation}
in the Epstein regime, with $a_\mathrm{gr}$ the radius of a single (pebble or dust) grain and $\rho_\mathrm{gr}$ the density of the solid material (i.e., silicate). The radial velocity of the solids is given by
\begin{equation}
    u_\mathrm{solid} = \frac{u_\mathrm{gas}}{\St^2 + 1} - \frac{2 u_\mathrm{grad}}{\St + \St^{-1}}
\end{equation}
where
\begin{equation}
    u_\mathrm{gas} = -\frac{3}{R^{1/2} \Sigma_\mathrm{gas}} \frac{\partial}{\partial R} \left(R^{1/2} \nu \Sigma_\mathrm{gas}\right)
\end{equation}
is the gas velocity and 
\begin{equation}
    u_\mathrm{grad} = -\frac{E_d}{2 \rho_\mathrm{gas} \Omega} \frac{\partial p_\mathrm{gas}}{\partial R}
    \label{eqn:ugrad}
\end{equation}
is the velocity due to the gas pressure gradient. $E_d$ is a drift efficiency parameter and $p_\mathrm{gas} = \rho_\mathrm{gas} c_s^2$ is the gas pressure. \citet{Birnstiel+2010A&A} gives more detail on these equations.

Again, we must make some assumption about the vertical distribution of solids to determine $\rho_\mathrm{solid}$. We make the same vertical Gaussian assumption as for the gas, but, to simulate settling, we allow the scale height of the pebbles to be a fraction $\xi_\mathrm{pebbles}$ of the gas scale height, so $h_\mathrm{pebbles} = \xi_\mathrm{pebbles} h_\mathrm{gas}$. We use $\xi_\mathrm{dust} = 1$ such that the dust is not settled. \revisionone{For the pebbles, we take $\xi_\mathrm{pebbles} = 0.1$.}

\subsection{Adsorption and desorption}

Finally, adsorption, the process by which atoms and molecules stick to a solid surface, and desorption, in which the atoms and molecules leave the surface, must be included as source terms. \citet{Hollenbach+2009ApJ} gives the adsorption timescale as
\begin{equation}
    \tau_\mathrm{ads} = \left(n_\mathrm{solid} \sigma_\mathrm{gr} v_\mathrm{therm}\right)^{-1}
\end{equation}
where $n_\mathrm{solid}$ is the local number density of solids, $\sigma_\mathrm{gr} = \pi a_\mathrm{gr}^2$ is the cross-sectional area of a single grain, and $v_\mathrm{therm} = \sqrt{8 k_B T / \pi m}$ is the thermal velocity of the atom or molecule of interest. Inverting the timescale, we find the adsorption rate
\begin{equation}
    R_\mathrm{ads} = n_\mathrm{solid} \sigma_\mathrm{gr} v_\mathrm{therm}
    \label{eqn:adsrate}
\end{equation}
per atom or molecule.

For desorption, \citet{Hollenbach+2009ApJ} gives the rate per molecule of ice
\begin{equation}
    R_\mathrm{des} = \nu_\mathrm{att} \exp\!\left(-\frac{T_\mathrm{bind}}{T}\right),
    \label{eqn:desrate}
\end{equation}
where $\nu_\mathrm{att}$ is the attempt frequency --- \revisionone{the vibrational frequency of the atoms and molecules on the surface} --- of order $10^{12}$~\s[-1], and $T_\mathrm{bind}$ is the binding energy of the species of interest (4800~\K\ for \ce{H2O} and 960~\K\ for \ce{CO}, \citealt{Aikawa+1996ApJ}).

Combining Equations~\ref{eqn:adsrate} and \ref{eqn:desrate}, we find the volumetric source terms
\begin{equation}
    s_\mathrm{gas} = R_\mathrm{des} n_\mathrm{solid} - R_\mathrm{ads} n_\mathrm{gas}.
\end{equation}
To find the appropriate source term for the surface density equations above, we must integrate $s_\mathrm{gas}$ vertically and multiply by the species' mass $m$. We find
\begin{equation}
    S_\mathrm{ads} = \frac{\sigma_\mathrm{gr} v_\mathrm{therm} \Sigma_\mathrm{gas} \Sigma_\mathrm{solid}}{\sqrt{2 \pi} m_\mathrm{gr} h_\mathrm{gas} \sqrt{1 + \xi_\mathrm{solid}^2}}.
    \label{eqn:Sads}
\end{equation}
and
\begin{equation}
    S_\mathrm{des} = m \int \limits_{-\infty}^{\infty} R_\mathrm{des} n_\mathrm{solid} \diff z = R_\mathrm{des} \Sigma_\mathrm{solid}.
\end{equation}
(Equation~\ref{eqn:Sads} is derived in Appendix~\ref{adx:sourceterms}.) \revisionone{These source terms both have units of \g~\cm[-2]~\s[-1]\ and represent the rates at which the surface density changes due to adsorption and desorption processes, respectively.}

Thus, the source terms for surface density equations are given by
\begin{equation}
    S_\mathrm{gas} = S_\mathrm{des} - S_\mathrm{ads}
    \label{eqn:gassrc}
\end{equation}
and
\begin{equation}
    S_\mathrm{solid} = S_\mathrm{ads} - S_\mathrm{des}.
    \label{eqn:dustsrc}
\end{equation}
\revisionone{These source terms encode the rate at which the surface densities of gas and solid species are changing due to the adsorption and desorption chemistry in our model.}

\subsection{Temperature structure}
\label{sec:temperature}

The temperature field presents a challenge by itself. Temperature appears in Equation~\ref{eqn:pringle} through the viscosity term, and so it contributes to the gas dynamics. Yet the gas dynamics play a role in determining the dust dynamics, which, through radiative transfer from the central star, determine the temperature. In addition, the intrinsic luminosity of the star is expected to change significantly over the timescales presented here \citep{Siess+2000A&A}. One way to solve this circular problem is through iteration, as in \citet{Price+2020ApJ}. However, that procedure would be more computationally costly when coupled to the code we have described here.

Instead of seeking a self-consistent solution, as in \citet{Price+2020ApJ}, we follow a simpler procedure to capture the approximate temperature structure. Noting that the bulk surface density, and therefore dust grain surface density, does not change significantly over time, we use RADMC-3D version 0.41 \citep{radmc3d} to compute a temperature structure with a self-similar dust initial condition (i.e., same form as Equation~\ref{eqn:selfsimilar}), assuming a dust-to-gas ratio of $0.01$ and using the \revisionone{interpolated DSHARP opacities \citep{dsharpopac}}. 
We note that this \revisionone{procedure} is an approximation, since we are not taking into account the evolving dust and pebble surface density, but it provides sufficient accuracy for our proof-of-concept purposes.

Next, we fit a power law $T \propto R^{-\beta}$ to the output from RADMC-3D, \revisionone{limited to the region between 2~\au\ and 20~\au\ to avoid edge effects and unphysical behavior far from the star. Though we run RADMC-3D with two dust populations, the temperatures are virtually equal, so we assume a power law slope of $-0.41$ and appropriate intercept parameter, which reasonably captures the behavior of both populations, and use that same power law for both when solving the differential equations.}

To take into account a changing stellar luminosity over time, we use the \citet{Siess+2000A&A} web server to compute stellar radii $R_\star$ and effective temperatures $T_\mathrm{eff}$ over the lifetime of the disk. Then, inspired by \citet{ChiangGoldreich1997ApJ}, Equation~12, we see that the disk temperature scales by a factor $f \propto R_\star^{1/2} T_\star$. We compute this factor from the isochrons and scale it by the initial value such that $f \leq 1$ at all times, i.e., the disk temperature is decreasing over time, primarily due to radial contraction decreasing the bolometric luminosity of the central star.

\begin{figure}
    \centering
    \includegraphics[width=0.95\linewidth,keepaspectratio]{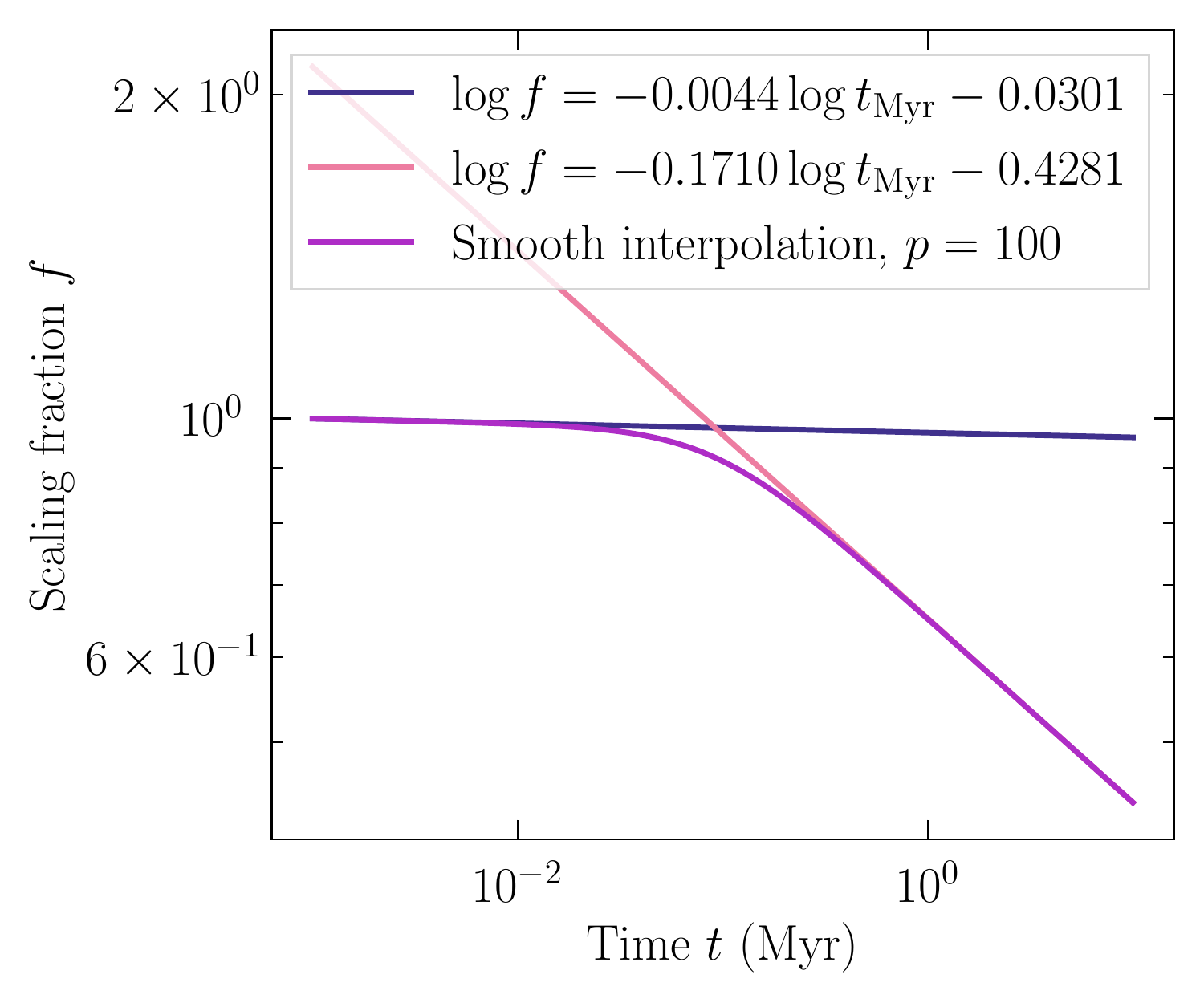}
    \caption{Temperature scaling fraction as determined by fitting \citet{Siess+2000A&A} isochrons with two power laws and then interpolating smoothly between them. The complete procedure is described in Section~\ref{sec:temperature}. The parameters of the lines are given in the legend, and the interpolation ``power'' $p$ is chosen to give a smooth curve to the intersection of the lines.}
    \label{fig:fraction}
\end{figure}

\begin{figure}
    \centering
    \includegraphics[width=0.9\linewidth,keepaspectratio]{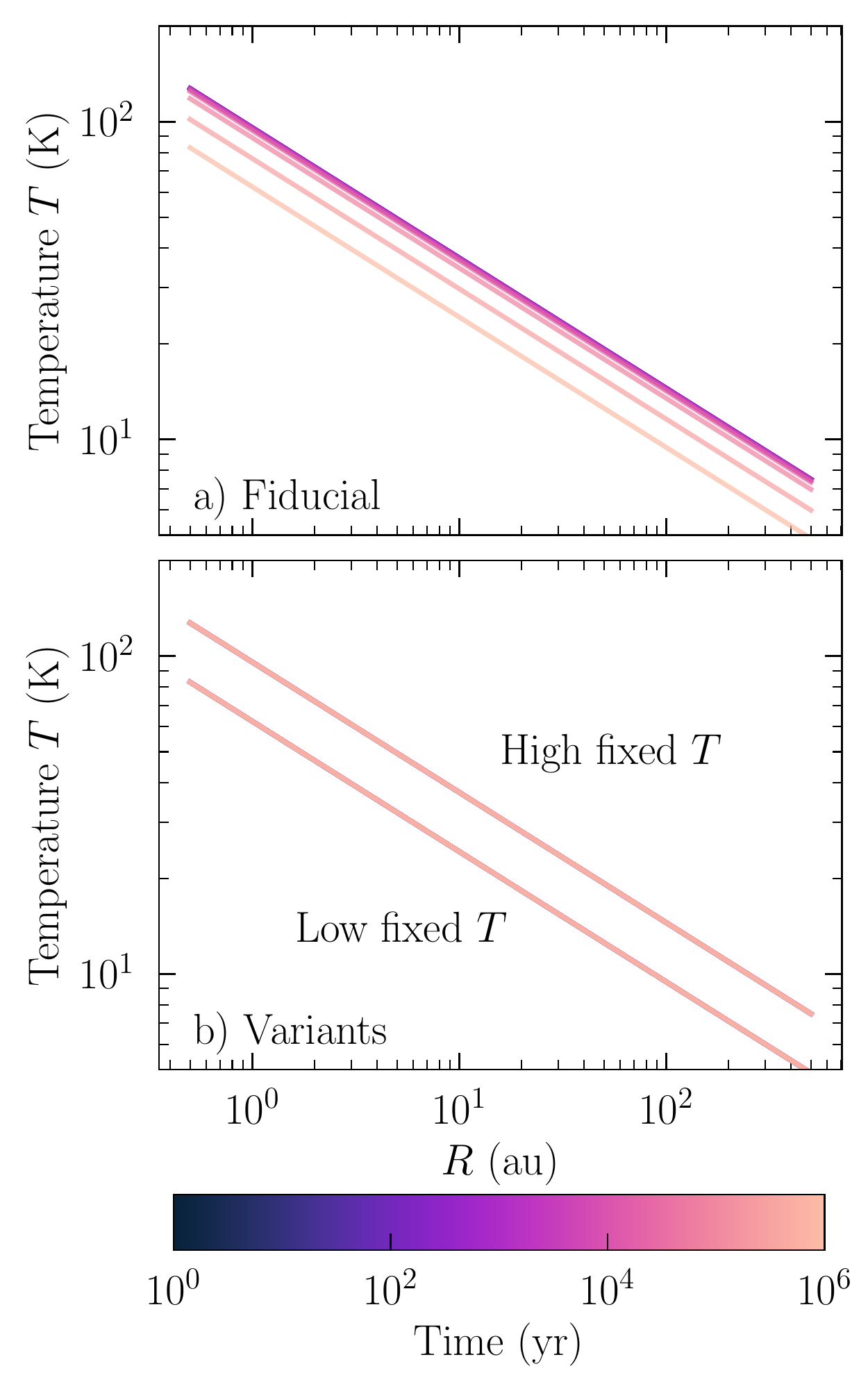}
    \caption{In panel (a), we show the temporal and spatial evolution of the disk temperature (assumed vertically invariant for the purposes of the model) for the fiducial case. In panel (b), we show the two variants explored: a high temperature case (upper) and low temperature case (lower), both of which are held fixed in time. \revisionone{The line color in all panels indicates logarithmically-increasing time.}}
    \label{fig:temperature}
\end{figure}

Finally, we perform a fit to the two regimes we observe in $f$ --- a flat, early-time regime and a sloped, late-time regime --- and join the two regimes by smooth interpolation. This interpolation takes the same form as Equation~\ref{eqn:surfdens}, but with a parameter $p = 100$ that is more appropriate for this data. See Figure~\ref{fig:fraction} for the parameters in each regime and the final interpolation. Figure~\ref{fig:temperature} shows the resulting temperature that is used in the fiducial model alongside two fixed-temperature models representing the beginning and end state.

\subsection{Solution procedure}

To solve Equations~\ref{eqn:pringle} and \ref{eqn:birnstiel} with source terms given by Equations~\ref{eqn:gassrc} and \ref{eqn:dustsrc}, we require approximations of first and second derivatives in radius. We use a logarithmically-spaced mesh in $R$ and second-order accurate finite difference derivatives estimated with Equations~\ref{eqn:firstderiv} and \ref{eqn:secondderiv}. Where appropriate, we switch to first-order accurate upwind finite difference derivatives.

To advance the solution in time, we use the backward differentiation formula (BDF) implementation in the Portable, Extensible Toolkit for Scientific Computation (PETSc) \citep{petsc-efficient,petsc-web-page,petsc-user-ref} time stepping (TS) \citep{petsc-ts} module. We use the PETSc internal colored finite difference Jacobian and solve the resulting linear system with the MUltifrontal Massively Parallel sparse direct Solver (MUMPS) \citep{mumps1,mumps2}.

The system of partial differential equations we finally solve is in nine quantities. The bulk gas, pebble, and dust densities are treated according to Equations~\ref{eqn:pringle} and \ref{eqn:birnstiel} with no source terms. Then, we consider \ce{H2O} and \ce{CO} in gas, as ice on pebbles, and as ice on dust grains by adding the appropriate source term to the right-hand sides of Equations~\ref{eqn:pringle} and \ref{eqn:birnstiel}. We evolve the equations to $1$~\Myr\ on the spatial domain $\left[0.5~\au, 500~\au\right]$.

\section{Results}
\label{sec:results}

\begin{deluxetable*}{lcccc}
    \tablecaption{Various model cases and parameter values. \label{tbl:models}}
    \tablehead{\colhead{Identifier} & \colhead{Viscosity parameter $\alpha$} & \colhead{Drift efficiency $E_d$} & \colhead{Initial \ce{CO}/\ce{H2O}} & \colhead{Temperature model}}
    \startdata
    Fiducial & $10^{-3}$ & $0.1$ & $20\%$ & time-evolving \\
    Low $\alpha$ & $10^{-4}$ & $0.1$ & $20\%$ & time-evolving \\
    Low drift & $10^{-3}$ & $0.01$ & $20\%$ & time-evolving \\
    High drift & $10^{-3}$ & $0.9$ & $20\%$ & time-evolving \\
    Low \ce{CO} & $10^{-3}$ & $0.1$ & $1\%$ & time-evolving \\
    High \ce{CO} & $10^{-3}$ & $0.1$ & $100\%$ & time-evolving \\
    Low fixed $T$ & $10^{-3}$ & $0.1$ & $20\%$ & static, $t = 1~\Myr$ \\
    High fixed $T$ & $10^{-3}$ & $0.1$ & $20\%$ & static, $t = 0~\Myr$ \\
    \enddata
\end{deluxetable*}

\subsection{Fiducial model}


\begin{figure*}
    \centering
    \includegraphics[width=0.9\linewidth,keepaspectratio]{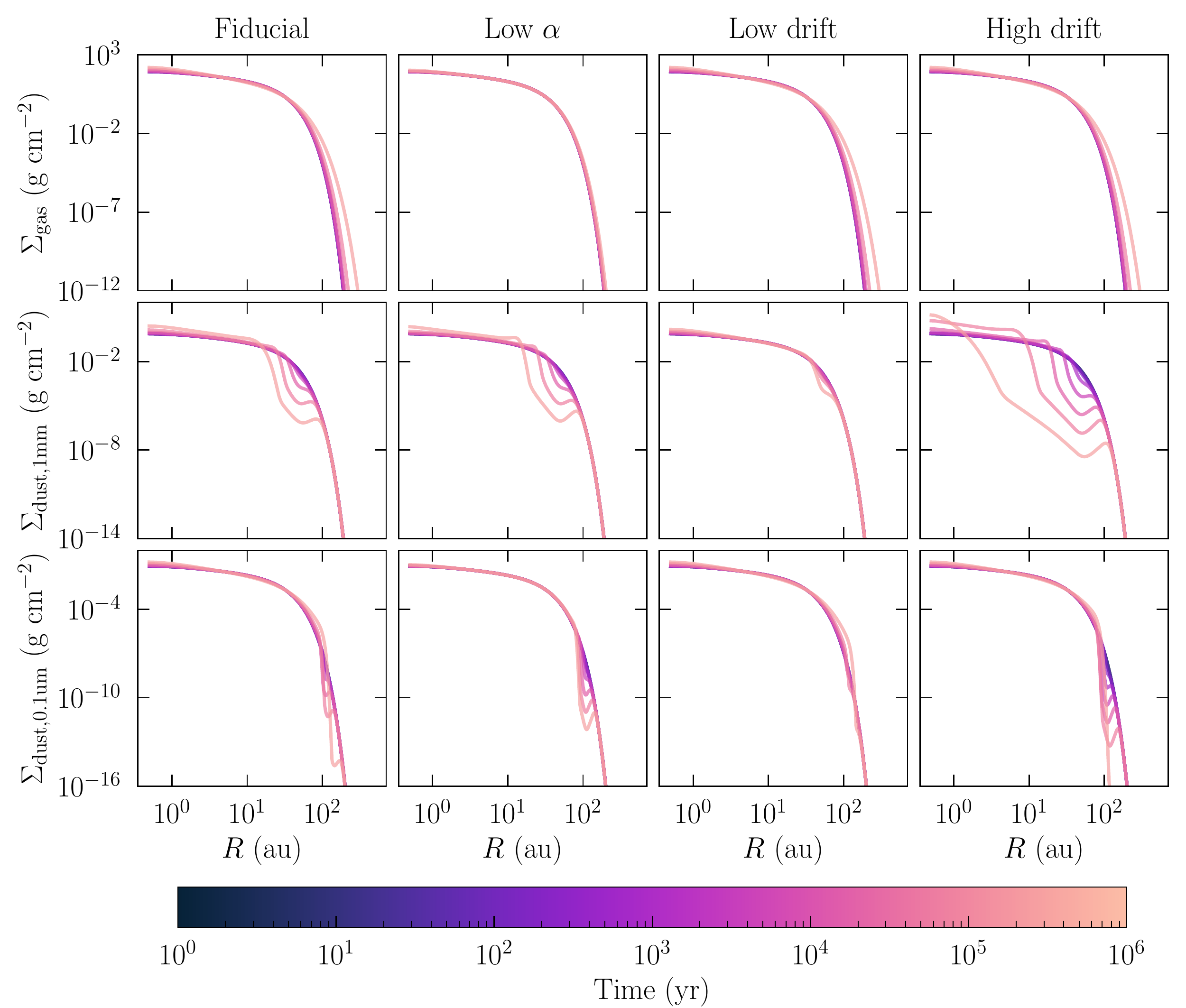}
    \caption{Temporal and spatial evolution of the three bulk surface densities for four selected model cases. The rows represent the surface densities of each type (gas, 1~\mm\ pebbles, and 1~\um\ dust, from top to bottom, respectively) while the columns represent the different models. Temporal evolution is shown by the color gradient, which extends through time on a logarithmic scale from darker to lighter colors. The most notable feature is the development of a density deficit in the solids around 100~\au\ caused by drift; this is observed in both pebbles and dust, but the pebbles exhibit a stronger effect because they drift more efficiently than the dust, which is well-coupled to the gas.}
    \label{fig:model}
\end{figure*}

\begin{figure*}
    \centering
    \includegraphics[width=0.88\linewidth,keepaspectratio]{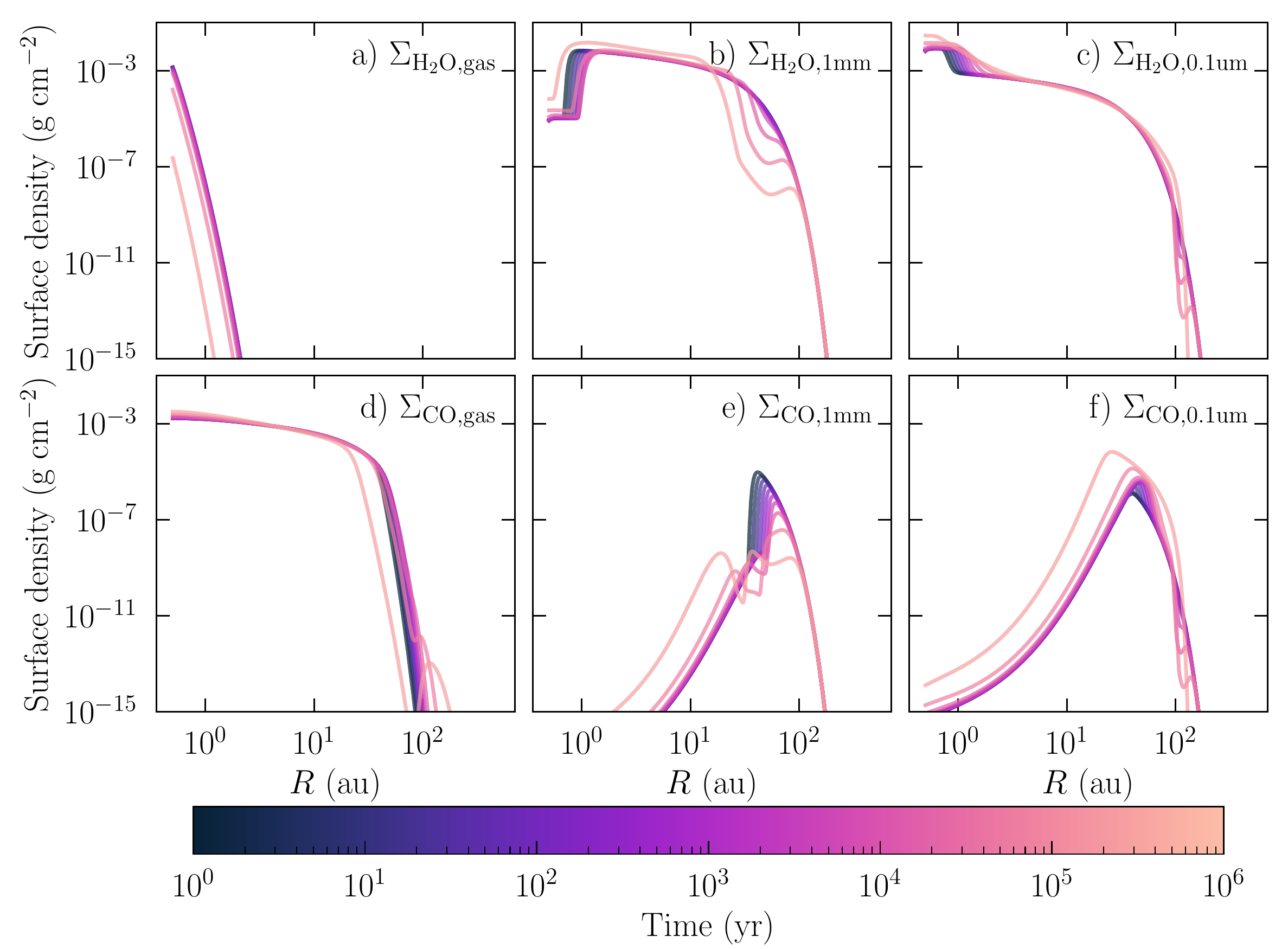}
    \caption{Fiducial model evolution of the surface densities of \ce{H2O} (top row) and \ce{CO} (bottom row) in the gas phase (left column) and solid phases (middle and right column). The pebble deficit from Figure~\ref{fig:model} is echoed in the water ice, but the \ce{CO} experiences a very different behavior than the bulk solids.}
    \label{fig:chemistry}
\end{figure*}

In Figure~\ref{fig:model} (first column), we show the behavior of the bulk gas, pebbles, and dust over time and radius in the fiducial model. While the gas behavior shows simple viscous spreading, the pebbles and dust show more interesting behavior. The 1~\mm\ pebbles form a shallow gap-like structure at about 30--100~\au. This position coincides with the radius where the Stokes number goes to unity, and thus where the pebbles move fastest. As a result, at smaller radii, the pebbles move inward, and, at larger radii, the pebbles move outward, resulting in a pebble deficit at about 100~\au. The 0.1~\um\ dust forms a similar structure at larger radii.

Figure~\ref{fig:chemistry} shows that the ice-coated solids do not universally follow the same trends as the bulk. While \ce{H2O} on pebbles and dust forms the gap-like structure near 100~\au, there is a second pebble and dust deficit at 1~\au, the \ce{H2O} snowline, where there is also a rapid increase in \ce{H2O} vapor surface density. The behavior of \ce{CO} is significantly different from that of \ce{H2O}. The gap at $30$--$100$~\au\ is much shallower, and only clearly visible at late times. Analogous to \ce{H2O}, there is a rapid drop in \ce{CO} dust and pebble surface density at the CO snowline. Figure~\ref{fig:chemistry} already shows a clear change in the \ce{CO}/\ce{H2O} surface density ratio in the outer disk.


In Figure~\ref{fig:multipanel}, we present the main results of this paper both for the fiducial model and for a small parameter study (see next section). Each panel in the figure shows the evolution of the \ce{CO}/\ce{H2O} ratio in two ways: On the left, we show the variation over time and space on the vertical and horizontal axes, respectively. On the right, in a smaller panel, we integrate over radius in the region shown and show the evolution of the ice mass --- total and where \ce{CO}/\ce{H2O} $\geq 1$ --- over time. In Figure~\ref{fig:multipanel}a, we show the predicted ratio of \ce{CO}/\ce{H2O} for our fiducial model, and we find that a maximum \revisionone{ratio near unity} is achieved by 1~\Myr\ in the region between about $20$ and $200$~\au, and that this feature takes the shape of a funnel when observed in the space-time plane. This enhanced material accounts for \revisionone{an average of $40\%$} of the disk mass. See Table~\ref{tbl:outcomes} for similar measurements of each model case that follows.

\subsection{Parameter study}


While our fiducial model results are encouraging in explaining anomalous, \ce{CO}-enhanced comets, we also seek to understand the robustness of this result to changes in disk parameters, relatively unconstrained by observations or detailed simulations. The first parameters of interest are the viscosity parameter and drift efficiency. The viscosity parameter $\alpha$ ultimately sets the diffusion coefficient, which directly controls the gas's diffusion and the diffusive flux of the solids. Figure \ref{fig:model} (second column) shows that reducing this parameter only has some minor effects on the dust and pebble evolution. The drift efficiency $E_d$ influences the coupling of solids to gas but only appears in the dust velocity, and therefore leaves gas motion unchanged. Figure \ref{fig:model} (third and fourth columns) show that increasing and decreasing this parameter dramatically changes the drift and therefore depletion of solids in the outer disk regions.

\begin{figure*}
    \centering
    \includegraphics[width=0.85\textwidth, keepaspectratio]{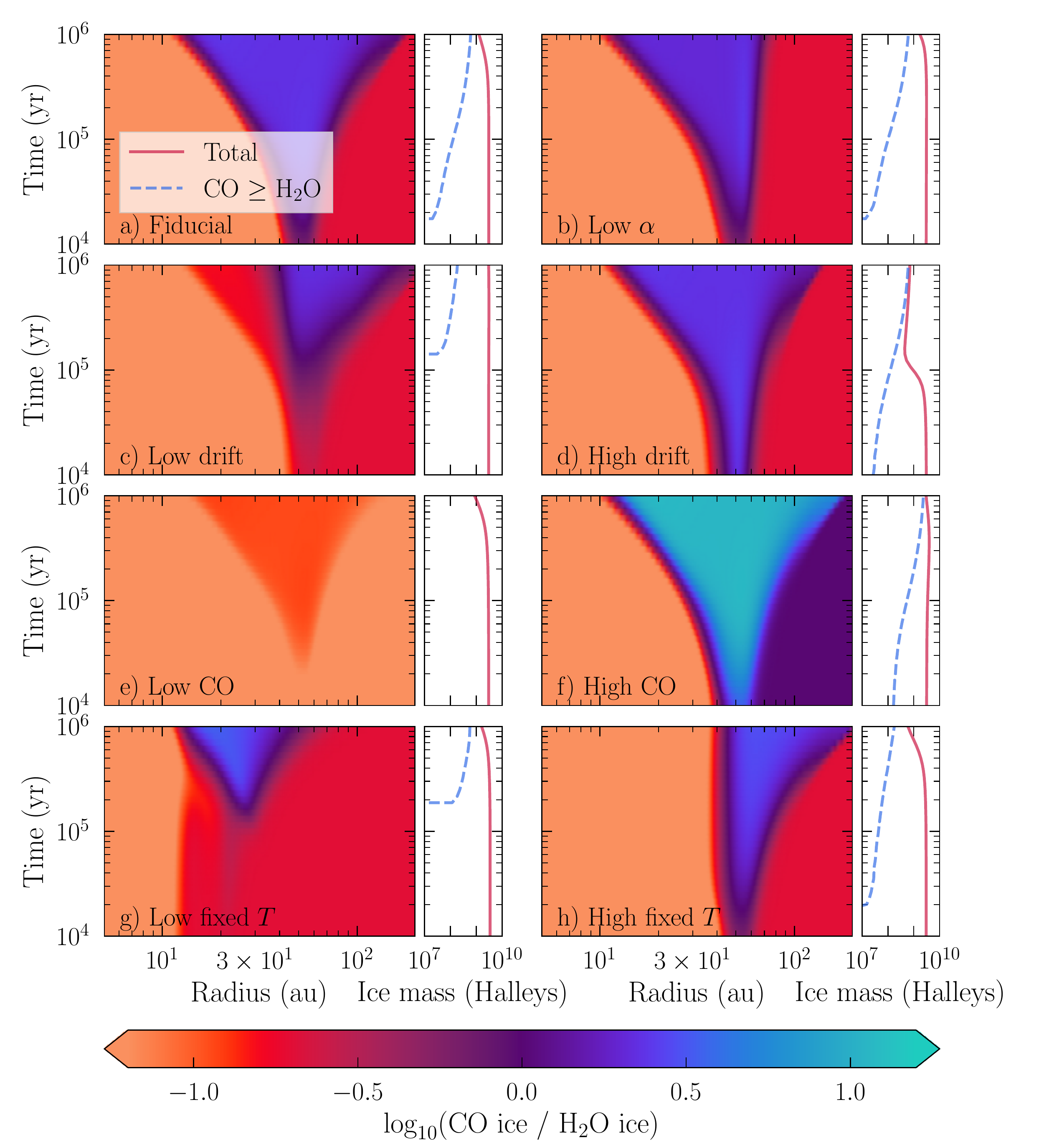}
    \caption{Evolution of the \ce{CO}/\ce{H2O} ice ratio as a function of time and space in the disk model. Each large panel represents a different set of conditions or parameters used, which can be found listed in Table~\ref{tbl:models}. In each large panel, the two smaller panels show the evolution of the ice ratio (left, two-dimensional color plot) and the total mass in enhanced ice over time (right, line plot). In the line plot, the ice mass is measured in units of Halley's comet's mass, to give the reader an idea of how many comets could be formed from the material if formation was 100\% efficient.}
    \label{fig:multipanel}
\end{figure*}

Figure~\ref{fig:multipanel}b shows the enhancement of the \ce{CO}/\ce{H2O} ratio in ice for a model with $\alpha$ reduced by a factor of ten compared to the fiducial model. Reducing $\alpha$ makes the viscosity smaller everywhere, which, in turn, amounts to making the diffusion coefficient smaller. Thus, we would expect that disk material would experience less viscous spreading in this case, and, indeed, we see that the characteristic ``funnel'' shape of the \ce{CO}-enhanced region in time and space is truncated and does not reach 100~\au, while drift still carries material inward towards the star. The amount of \ce{CO}-\revisionone{enhanced} ice is modest, but it is certainly present.

In Figures~\ref{fig:multipanel}c and \ref{fig:multipanel}d, we show the enhancement of the \ce{CO}/\ce{H2O} ratio in ice for the low and high drift models, respectively; for these test cases, we fixed $E_d$ at $0.01$ and $0.9$, changing the efficiency of the coupling to the gas pressure derivative. \revisionone{The models achieve about the same maximum \ce{CO}/\ce{H2O} ratio, but with very different fractions of \ce{CO}-\revisionone{enhanced} ice; i.e., the low-drift enhancement feature is visibly smaller in the space-time plane.} We immediately see, then, that the efficiency of the radial drift of pebbles and dust is very important for predicting the amount of mass available for making comets like 2I/Borisov and C/2016 R2 (PanSTARRS). We return to this in the discussion section.

The third parameter of interest is the initial \ce{CO}/\ce{H2O} ratio. We test two possibilities in addition to the fiducial model: A high \ce{CO}/\ce{H2O} value of $100\%$ and a low \ce{CO}/\ce{H2O} value of $1\%$. We choose these end-member cases because, \revisionone{while typical comets have \ce{CO}/\ce{H2O} of about $4\%$ \citep{BockeleeMorvan&Biver2017}, they may have $1\%$ or less \citep{MummaCharnley2011AnnuRev}}, while the interstellar medium has up to $100\%$ with a large errorbar \citep{Oberg2016ChemRev}. Note that the amount of \ce{CO} has no effect on the bulk dynamics; it only affects the chemical evolution of the disk. In Figures~\ref{fig:multipanel}e and \ref{fig:multipanel}f, we show the chemical evolution of the disk in these two cases. We find that the low \ce{CO} model is \revisionone{not} able to reach a \ce{CO}/\ce{H2O} ratio of unity; see Figure~\ref{fig:multipanel}e. On the other hand, the high \ce{CO} model easily reaches values of \ce{CO}/\ce{H2O} $\gtrsim 10$ by about $1$~\Myr, as shown in Figure~\ref{fig:multipanel}f, and this material accounts for a large fraction of the total disk mass.

The effect of a static, low temperature profile and a static, high temperature profile on our disk model is shown in Figures~\ref{fig:multipanel}g and \ref{fig:multipanel}h, respectively. For these cases, we artificially fixed the temperature at its final or initial value, as appropriate (recall that the temperature strictly decreases with time, and see Figure~\ref{fig:temperature} for the radially-dependent structures we adopted). These models reach roughly the same level of \ce{CO}/\ce{H2O} ice enhancement, but the radii where the enhancement occurs are shifted. In the low temperature model, the onset of the enhancement is delayed in time. The high temperature model's enhanced region is shifted to larger radii because the disk is warmer everywhere, and so ice will desorb off the grains in this model farther out than in the fiducial model. Most importantly, the details of the temperature structure and evolution are not critical for the formation of a substantial amount of \ce{CO} ice; \revisionone{both static temperature models achieve at least 30\% \ce{CO}-enhanced ice (see Table~\ref{tbl:outcomes})}.

Finally, we summarize our results numerically in Table~\ref{tbl:outcomes} in terms of maximum \ce{CO}/\ce{H2O} ratio, total number of \ce{CO}-\revisionone{enhanced} Halley-mass comets, and the mass fraction of the ice in the disk that is CO-\revisionone{enhanced} by the end of the simulation. Most models achieve a \ce{CO}/\ce{H2O} ratio above \revisionone{unity} (only the low CO model achieves a lower maximum ratio). The lowest ratio (low initial \ce{CO} model) is just over $0.1$, and the largest ratio (high initial \ce{CO} model) is greater than $10$, revealing a \revisionone{monotonic} dependence on \ce{CO} initial abundances. The low drift model produces the next-least amount of \ce{CO}-enhanced ice. The fraction of ice in the region $\left[5~\au, 200~\au\right]$ that is \ce{CO}-enhanced is, on average, about $40\%$, but in the high-drift model it is all of $87\%$, indicating that most water ice has been lost from the system due to pebble drift. The maximum number of \ce{CO}-\revisionone{enhanced} Halley-like comets that could be formed in the disks is between $10^8$ and $10^{10}$, though this assumes a formation efficiency of 100\% from the dust and pebbles and no additional mixing, trapping, or drift. While a large range of values are possible, we emphasize that there is \revisionone{almost} always a region where there is \revisionone{some} \ce{CO} ice enhancement relative to \ce{H2O} ice.

\begin{deluxetable*}{lcccc}
    \tablecaption{Model outcomes. Masses are measured over the same region shown in Figure~\ref{fig:multipanel}. \label{tbl:outcomes}}
    \tablehead{\colhead{Identifier} & \colhead{Highest \ce{CO}/\ce{H2O} ratio} & \colhead{Total \ce{CO}-\revisionone{enhanced} Halley-mass comets} & \colhead{Fraction of \ce{CO}-\revisionone{enhanced} ice\tablenotemark{a}}}
    \startdata
    Fiducial & $2.39$ & $\expn{6}{9}$ & $48\%$ \\
Low $\alpha$ & $2.43$ & $\expn{6}{9}$ & $36\%$ \\
Low drift & $2.21$ & $\expn{2}{8}$ & $6\%$ \\
High drift & $2.56$ & $\expn{6}{9}$ & $87\%$ \\
Low CO & $0.12$ & -- & $0\%$ \\
High CO & $11.93$ & $\expn{2}{9}$ & $78\%$ \\
Low fixed $T$ & $3.47$ & $\expn{6}{9}$ & $35\%$ \\
High fixed $T$ & $2.96$ & $\expn{2}{8}$ & $30\%$ \\
    \enddata
    \tablenotetext{a}{\revisionone{We define \ce{CO}-enhanced as ice with $\Sigma_\mathrm{CO} \geq \Sigma_{\mathrm{H}_2\mathrm{O}}$.}}
\end{deluxetable*}

\section{Discussion}
\label{sec:discussion}


We have explored the role of dust drift in changing the local ice composition in a protoplanetary disk midplane. Using models that assume simple ices composed of \ce{CO} and \ce{H2O} and allowing for adsorption and desorption, we find that parameters controlling the dynamics, such as the drift efficiency and viscosity, as well as the temperature (which affects dynamics indirectly), all play an important role in determining the specific amount of \ce{CO} enhancement relative to \ce{H2O} as well as the distribution of ices by 1~\Myr. Yet, across our models, there is consistently a region of our disk that displays \ce{CO}/\ce{H2O} ice enhancement compared to the initial \ce{CO}/\ce{H2O} abundance ratio, independent of the choice of parameters, in all cases we have explored.


Why does this enhancement occur? Most of the water in our model is in the form of ice. Drift carries the water ice-laden pebbles and dust inward, creating an ice deficit. We can see from Figure~\ref{fig:chemistry} that the water ice deficit forms around 100~\au\ and spreads out in time. Meanwhile, even though we start with \ce{CO} as ice, interior to its snowline, it initially sublimates quickly. The gas-phase \ce{CO} crosses the snowline as it viscously spreads out. This ``new'' \ce{CO} enters the water ice deficit region and then freezes out onto whatever solids remain. In Figure~\ref{fig:chemistry}c and \ref{fig:chemistry}f, we see that, while the amount of \ce{CO} on pebbles decreases over time, the amount of \ce{CO} on dust increases. The radial process we have described is similar to the ``vertical cold finger effect'' described by \citet{Meijerink+2009ApJ}, where water is depleted in the upper disk layers because of diffusive transport and settling. In addition, this work is consistent with the results of \citet{RosJohansen2013A&A}, which found significant solid enhancement caused by transport across the radial snowline. This work demonstrates that there is likely to be a complex interplay with the evolution of solids and the chemical composition of the ice mantles they harbor. Future work should explore these connections with more advanced chemistry along with ice chemistry and/or isotopic chemistry, to fully understand the relationship between grain drift, viscous spreading across snowlines, and the resulting chemistry.



\revisionone{While we have limited ourselves in this paper to only two grain sizes, a more realistic simulation would use a continuous distribution of grain sizes. We expect that the largest grain size is the driving factor of the location of the inner edge of the enhancement feature. When the largest size is reduced from 1~\mm, the largest size we considered here, drift becomes less efficient; when it is increased, drift becomes more efficient. Drift greatly influences the location of the inner edge of the ``funnel'' we observe in the models we present here. Since the mechanism proposed above only needs some small grain population to be entrained with the gas and some large population that drifts efficiently, we theorize that the exact distribution of grain sizes does not strongly influence our results.}

\section{Conclusions}
\label{sec:conclusions}

We present models of the surface density evolution of a viscously-evolving protoplanetary disk, including the effect of grain drift, with the goal of explaining the observations of \ce{CO}-enriched comets. To explore how midplane \ce{CO} and \ce{H2O} abundances in gas and ice evolve within this dynamic framework, we include simple adsorption and desorption chemistry to capture the interplay of dust transport and snowlines. We find that \revisionone{most of} our disk models readily produce a region where \ce{CO} ice is more abundant than \ce{H2O} ice. These results indicate that forming \ce{CO}-enriched comets may not be so unusual.

\revisionone{On the other hand, the fact remains that we have not observed very many \ce{CO}-enriched comets to date. Assuming our Solar System originated with a nominal amount of \ce{CO}, there may be some selection bias that causes \ce{CO}-poor comets to be observed more frequently.}

\citet{Fitzsimmons+2019ApJL} and \citet{Xing+2020ApJL} conclude that the extrasolar comet 2I/Borisov is in most ways --- excluding its high \ce{CO}/\ce{H2O} ratio --- similar to Solar System comets. Our results support the conclusion that the \ce{CO}/\ce{H2O} ice enhancement commonly occurs in the outer disk for solar-type stars, between $20$ and $100$~\au. Perhaps comets that form so far out are more easily ejected due to being weakly gravitationally bound to their host star. 2I/Borisov may be an example of this mechanism at work. While dynamical simulations are beyond the scope of the present work, it would be interesting to compare the expected distribution of formation locations of extrasolar comets pre-ejection with the chemical patterns found here, to further test this hypothesis. 

\clearpage
\acknowledgements

\vbox{E.M.P. gratefully acknowledges support from National Science Foundation Graduate Research Fellowship Program (GRFP) grants DGE1144152 and DGE1745303 and helpful conversations with Prof. Zhaohuan Zhu and Prof. Paul C. Duffell. This work was supported by an award from the Simons Foundation (SCOL \# 321183, KO).  L.I.C. gratefully acknowledges support from the David and Lucille Packard Foundation, the VSGC New Investigators Award, the Johnson \& Johnson WiSTEM2D Award, and NSF AST-1910106.

This work was inspired by conversations at the May 2019 ``ExoComets: Understanding the Composition of Planetary Building Blocks'' workshop, and we are grateful for the Lorentz Center's support of this event.}

\software{RADMC-3D version 0.41 \citep{radmc3d}, PETSc \citep{petsc-efficient,petsc-web-page,petsc-user-ref,petsc-ts}, MUMPS \citep{mumps1,mumps2}, Siess isochrons \citep{Siess+2000A&A}, CMasher \citep{cmasher}}

\bibliographystyle{aasjournal}
\bibliography{biblio}

\appendix

\section{Supplementary equations}

\subsection{Vertically-integrated source term}
\label{adx:sourceterms}

Since the evolution equations given in this paper are in terms of surface density, which is a vertically-integrated quantity, it is important to additionally vertically integrate the usual adsorption source term, as
\begin{align}
    S_\mathrm{ads} = m \int\limits_{-\infty}^{\infty} R_\mathrm{ads} n_\mathrm{gas} \diff z &= \sigma_\mathrm{gr} v_\mathrm{therm} \int\limits_{-\infty}^{\infty} \frac{\Sigma_\mathrm{dust}}{\sqrt{2 \pi} h_\mathrm{dust} m_\mathrm{gr}} \exp\!\left[-\frac{1}{2} \left(\frac{z}{h_\mathrm{dust}}\right)^2\right] \frac{\Sigma_\mathrm{gas}}{\sqrt{2 \pi} h_\mathrm{gas}} \exp\!\left[-\frac{1}{2} \left(\frac{z}{h_\mathrm{gas}}\right)^2\right] \diff z \\
    &= \frac{\sigma_\mathrm{gr} v_\mathrm{therm} \Sigma_\mathrm{gas} \Sigma_\mathrm{dust}}{\sqrt{2 \pi} m_\mathrm{gr} h_\mathrm{gas} \sqrt{1 + \xi_\mathrm{dust}^2}}.
\end{align}
Note that we would have missed an important correction factor had we na\"ively multiplied $n_\mathrm{gas}$ and $n_\mathrm{dust}$ without taking into account the vertical integration.

\subsection{Finite difference approximations}
On a finite grid in $x$ with points $\{ x_i \}$, we use the modified finite difference formulae
\begin{equation}
    \frac{\partial f}{\partial x} \approx \left(\frac{-h_{i+1}}{h_{i-1} \left(h_{i-1} + h_{i+1}\right)}\right) f_{i-1} + \left(\frac{1}{h_{i-1}} - \frac{1}{h_{i+1}}\right) f_i + \left(\frac{h_{i-1}}{h_{i+1} \left(h_{i-1} + h_{i+1}\right)}\right) f_{i+1}
    \label{eqn:firstderiv}
\end{equation}
and
\begin{equation}
    \frac{\partial^2 f}{\partial x^2} \approx \left(\frac{2}{h_{i-1} \left(h_{i-1} + h_{i+1}\right)}\right) f_{i-1} + \left(\frac{-2}{h_{i-1} h_{i+1}}\right) f_i + \left(\frac{2}{h_{i+1} \left(h_{i-1} + h_{i+1}\right)}\right) f_{i+1},
    \label{eqn:secondderiv}
\end{equation}
where $h_{i-1} = x_i - x_{i-1}$ and $h_{i+1} = x_{i+1} - x_i$, and $\{ f_i \}$ are samples of a smooth function $f\!\left(x\right)$. These formulae are general and second-order accurate, and they apply to  any irregularly-spaced grid.
\end{document}